\documentclass[10pt]{article}
\usepackage{amsmath}
\usepackage{braket}
\usepackage{titlesec}
\usepackage{graphicx}
\usepackage{hyperref}
\usepackage[margin=1in]{geometry}
\begin{document}






\centerline{\bf
Quantum Unsupervised and Supervised Learning on Superconducting Processors}
\vspace*{0.035truein}

\centerline{\footnotesize
Abhijat Sarma, 
Rupak Chatterjee\footnote{Corresponding Author: Rupak.Chatterjee@Stevens.edu},
Kaitlin Gili,
and Ting Yu}
\vspace*{0.015truein}
\centerline{\footnotesize\it Center for Quantum Science and Engineering and Department of Physics,}
\baselineskip=10pt
\centerline{\footnotesize\it Stevens Institute of Technology, Hoboken, New Jersey 07030, U.S.A.}

\vspace*{0.21truein}

\abstract{
Machine learning algorithms perform well on identifying patterns in many different datasets due to their versatility. However, as one increases the size of the dataset, the computation time for training and using these statistical models grows quickly. Quantum computing offers a new paradigm which may have the ability to overcome these computational difficulties. Here, we propose a quantum analogue to K-means clustering, implement it on simulated superconducting qubits, and compare it to a previously developed quantum support vector machine. We find the algorithm's accuracy comparable to the classical K-means algorithm for clustering and classification problems, and find that it has asymptotic complexity $O(N^{3/2}K^{1/2}\log{P})$, where $N$ is the number of data points, $K$ is the number of clusters, and $P$ is the dimension of the data points, giving a significant speedup over the classical analogue.
}

\section{Introduction}
Machine learning offers solutions to several classes of problems intractable through conventional computing means. For example, solutions to classification problems and regression of large datasets based on machine learning techniques are in general much more powerful than previously available solutions. These algorithms suffer in that they grow polynomial-wise with the size and dimension of the data, which leads to substantial run times when dealing with large datasets, coined "big data". The ability of data to be more efficiently stored and manipulated in quantum states has recently lead to the proposal of several quantum algorithms for machine learning \cite{var_circ1, var_circ2, varcircsim, qml, autoencoders, feedforward, stevens, featureHilbertSpace, circuitcentric, lloydqsvm}. In this paper, we develop a hybrid \textit{K}-means clustering algorithm in order to identify clusters in data and compare with similar purely classical algorithms. The algorithm relies on a distance measure, here taken to be Euclidean square distance. This distance can be calculated efficiently on a quantum computer, as we will show, in order to speed up the algorithm as a whole. Superconducting processors \cite{superc1, superc2, superc3}, such as those offered by IBMQ, present a quantum framework that is well-suited for circuit based algorithms, such as ours. Current superconducting quantum computers are too noisy to extract significant results, so we elected to test our algorithm on IBMQ's classical simulators which are designed to simulate quantum circuits on less noisy superconducting qubits. We first execute our quantum \textit{K}-means clustering algorithm on standard ad-hoc clustering datasets, and compare it to the classical \textit{K}-means algorithm. We then compare our quantum K-means clustering algorithm to the previously developed quantum SVM introduced in \cite{qsvm}, in terms of accuracy in solving trinary classification problems on real datasets. 

\section{Hybrid \textit{K}-means Clustering}

In machine learning theory, it is often mathematically convenient to consider the data as encoded in a vector. Therefore, each data point with \textit{P} different variables (features), can be encoded as a \textit{P}-dimensional feature vector. The total dataset is therefore a set of vectors in \textit{P}-dimensional space, known as input space. \textit{K}-means clustering is an unsupervised learning algorithm which considers the problem of partitioning \textit{N} feature vectors $\textbf{X}^{i}$ into \textit{K} \textit{subsets}, or \textit{clusters}. The algorithm seeks to find the \textit{K} clusters which minimize the dissimilarity between each cluster's members. Local minimization can be iteratively found with the standard \textit{K}-means algorithm, which relies on a distance measure, usually taken to be Euclidean square distance. However, it is worth mentioning that other distance measures can be very useful for certain problems in quantum physics \cite{quant_phase}. This algorithm is described below:
\begin{enumerate}
\item Randomly assign a cluster assignment \{1, 2, ..., \textit{K}\} to each of the observations $\textbf{X}^{i}$.
\item For each cluster $C_{k}$, calculate the \textit{vector of feature means} $(m_{k1}, m_{k2}, ..., m_{kP})$ where each feature mean $m_{kn}$ is the mean of the \textit{n}th feature of each vector in the \textit{k}th cluster. This vector is called the \textit{cluster centroid}.
\item Assign each observation $\textbf{X}^{i}$ to the cluster $C_{k}$ with the closest centroid according to the (Euclidean) distance measure.
\item Iterate between 2 and 3 until the assignments to clusters do not change.
\end{enumerate}

Sampling and estimating Euclidean distances between post-processed vectors on a classical computer is known to be exponentially hard. For big data sets, the algorithm becomes slow as convergence relies on repeated calculations of this distance measure. In the next section, we propose and implement a computationally cheap quantum algorithm for calculating the Euclidean distance. Note that although we use Euclidean distance in our clustering algorithm, one could use any distance measure.

Here we propose a hybrid algorithm which calculates cluster centroids and assigns features classically, but computes Euclidean (square) distance with a quantum circuit. To develop our quantum algorithm for estimation of Euclidean distance, we first must introduce the \textit{swap test} \cite{swap, swap1, swap2, swap3}. Consider a state \begin{equation}\ket{0}\otimes\ket{\psi}\otimes\ket{\varphi},\end{equation} consisting of an ancillary and two states needed for the overlap calculation. Perform a Hadamard transformation \textbf{H} on the ancillary followed by a \textbf{FREDKIN} gate (also known as a \textit{controlled swap}) on this state, \begin{equation}\begin{gathered}
\textbf{FREDKIN}(\textbf{H}\otimes\textbf{I}\otimes\textbf{I})\ket{0}\otimes\ket{\psi}\otimes\ket{\varphi}\\
=\bigl(\ket{0}\bra{0}\otimes\textbf{I}\otimes\textbf{I}+\ket{1}\braket{1}\otimes\textbf{SWAP}\bigr)\frac{1}{\sqrt{2}}[\ket{0}\otimes\ket{\psi}\otimes\ket{\varphi}+\ket{1}\otimes\ket{\psi}\otimes\ket{\varphi}]\\
=\frac{1}{\sqrt{2}}[\ket{0}\otimes\ket{\psi}\otimes\ket{\varphi}+\ket{1}\otimes\ket{\varphi}\otimes\ket{\psi}]\end{gathered}\end{equation}
Applying the Hadamard transformation once more to the ancillary qubit gives
\begin{equation}\begin{gathered}
\ket{\Psi}=(\textbf{H}\otimes\textbf{I}\otimes\textbf{I})\frac{1}{\sqrt{2}}[\ket{0}\otimes\ket{\psi}\otimes\ket{\varphi}+\ket{1}\otimes\ket{\varphi}\otimes\ket{\psi}]\\
=\frac{1}{2}[(\ket{0}+\ket{1})\otimes\ket{\psi}\otimes\ket{\varphi}+(\ket{0}-\ket{1})\otimes\ket{\varphi}\ket{\psi})]\\
=\frac{1}{2}\ket{0}\otimes[\ket{\psi}\otimes\ket{\varphi}+\ket{\varphi}\otimes\ket{\psi}]+\frac{1}{2}\ket{1}\otimes[\ket{\psi}\otimes\ket{\varphi}-\ket{\varphi}\otimes\ket{\psi}]
\end{gathered}\end{equation}
Finally, measure the state of the ancillary qubit. The probability of measuring $\ket{0}$, denoted by P(0), is given by 
\begin{equation}\begin{gathered}
\braket{\Psi|(\ket{0}\bra{0}\otimes\textbf{I}\otimes\textbf{I})|\Psi}=\frac{1}{4}\{\bra{\psi}\otimes\bra{\varphi}+\bra{\varphi}\otimes\bra{\psi}\}[\ket{\psi}\otimes\ket{\varphi}+\ket{\varphi}\otimes\ket{\psi}]\\
=\frac{1}{2}+\frac{1}{4}[(\bra{\psi}\otimes\bra{\varphi})(\ket{\varphi}\otimes\ket{\psi})+(\bra{\varphi}\otimes\bra{\psi})(\ket{\psi}\otimes\ket{\varphi})]\\
=\frac{1}{2}+\frac{1}{4}[\braket{\psi|\varphi}\braket{\varphi|\psi}+\braket{\varphi|\psi}\braket{\psi|\varphi}]
\end{gathered}\end{equation}
Therefore, as first shown in \cite{quantumworld},
\begin{equation}
P(0) = \braket{\Psi|(\ket{0}\bra{0}\otimes\textbf{I}\otimes\textbf{I})|\Psi} = \frac{1}{2}+\frac{1}{2}|\braket{\psi|\varphi}|^2
\end{equation}
The swap test therefore allows us to experimentally determine the overlap between two states $\ket{\psi}$ and $\ket{\varphi}$. This will be integral in calculating the Euclidean distance. 

The \textit{N} feature vectors of dimension \textit{P} can be written as $\textbf{X}^{i}=(X_{1}^{i}, X_{2}^{i}, ..., X_{P}^{i})$. In order to calculate Euclidean distance, we must first encode our feature vectors into Hilbert Space. Using the base-2 \textit{bit string configuration} $\ket{p}=\ket{p_{n-1}p_{n-2}...p_{1}p_{0}}, p=2^{0}p_{0}+2^{1}p_{1}+...2^{n-1}p_{n-1}, P=2^{n}$,
\begin{equation}
\ket{\textbf{X}^i} = \frac{1}{|\textbf{X}^i|}\sum_{p=1}^{P} X_{p}^{(i)}\ket{p}
\end{equation}
Note that the overlap between two of these states recovers the usual vector dot product, namely
\begin{equation}
\braket{\textbf{X}^{i}|\textbf{X}^{j}}=\frac{1}{|\textbf{X}^{i}||\textbf{X}^{j}|}\sum_{p=1}^{P}X_{p}^{(i)}X_{p}^{(j)}=\frac{1}{|\textbf{X}^{i}||\textbf{X}^{j}|}\textbf{X}^{i}\cdot\textbf{X}^{j}
\end{equation}
Next, we construct the following states
\begin{equation}\begin{gathered}
\ket{\psi}=\frac{1}{\sqrt{2}}[\ket{0}\otimes\ket{\textbf{X}^{i}}+\ket{1}\otimes\ket{\textbf{X}^{j}}]\\
\ket{\varphi}=\frac{1}{\sqrt{Z}}[|\textbf{X}^{i}|\ket{0} - |\textbf{X}^{j}|\ket{1}]
\end{gathered}\end{equation}
where $Z = |\textbf{X}^{i}|^2+|\textbf{X}^{j}|^2$ is a normalization constant. Performing a swap test between the first tensor product space of $\ket{\psi}$ and the state $\ket{\varphi}$ will give the overlap $|\braket{\psi|\varphi}|^2$ of the two states. To calculate this overlap, we first calculate the partial overlaps $\braket{\psi|\varphi}$ and $\braket{\varphi|\psi}$, which reduce to
\begin{equation}\begin{gathered}
\braket{\psi|\varphi} = \frac{1}{\sqrt{2Z}}[|\textbf{X}^{i}|\bra{\textbf{X}^{i}} - |\textbf{X}^{j}|\bra{\textbf{X}^{j}}]\\
\braket{\varphi|\psi} = \frac{1}{\sqrt{2Z}}[|\textbf{X}^{i}|\ket{\textbf{X}^{i}} - |\textbf{X}^{j}|\ket{\textbf{X}^{j}}]
\end{gathered}\end{equation}
Therefore, the complete overlap between these states $|\braket{\psi|\varphi}|^{2} = \braket{\psi|\varphi}\braket{\varphi|\psi}$ is
\begin{equation}\begin{gathered}
|\braket{\psi|\varphi}|^{2} = \braket{\psi|\varphi}\braket{\varphi|\psi}\\
=\frac{1}{2Z}\{|\textbf{X}^{i}|^2+|\textbf{X}^{j}|^2-|\textbf{X}^{i}||\textbf{X}^{j}|\braket{\textbf{X}^{i}|\textbf{X}^{j}}-|\textbf{X}^{j}||\textbf{X}^{i}|\braket{\textbf{X}^{j}|\textbf{X}^{i}}\}\\
=\frac{1}{2Z}\left\{|\textbf{X}^{i}|^2+|\textbf{X}^{j}|^2-2\textbf{X}^{i}\cdot\textbf{X}^{j}\right\}\\
=\frac{1}{2Z}\left\{|\textbf{X}^{i}-\textbf{X}^{j}|^2\right\}
\end{gathered}\end{equation}
The classical Euclidean distance is therefore proportional to the overlap of the states, specified in (10). The quantum algorithm for calculating the Euclidean distance is to perform a swap test between those two states such that
\begin{equation}\begin{gathered}
P(0) =  \frac{1}{2}+\frac{1}{2}|\braket{\psi|\varphi}|^2\\
= \frac{1}{2} + \frac{1}{4Z}|\textbf{X}^{i}-\textbf{X}^{j}|^2
\end{gathered}\end{equation}
or
\begin{equation}
|\textbf{X}^{i}-\textbf{X}^{j}|^2=Z(4P(0)-2).
\end{equation}
The state $\ket{\psi}$ can be easily constructed by simple Hadamard and CSWAP gates. The state $\ket{\varphi}$ can be created with a variation of quantum counting, followed by Grover's search algorithm, serving the dual purpose of estimating the normalization constant $Z$. 

Another recent quantum k-means clustering method exists, developed by Kerenidis et al \cite{qmeans}, though their methodology differs significantly from our own. They elect to create quantum states corresponding to each centroid, and use a distance estimation algorithm coupled with vector state tomography to update the centroid states in each iteration until the states are stable. Finally, a classical description of the final centroids is extracted by once again using vector state tomography. 

\section{Time Complexity Analysis}
In general, as the dimension and size of the data increases, we assert that the quantum distance measure will become less computationally expensive than the classical one to implement. For \textit{K}-means, we must calculate distance between each feature $\textbf{X}^{i}$ and each centroid $\textbf{C}^{k}=(m_{1}^{k}, m_{2}^{k}, ..., m_{P}^{k})$ in each iteration where $m_{p}^{k}=\frac{1}{|\textbf{C}^{k}|}\sum_{i\in \textbf{C}^{k}}{X_{p}^{i}}$ are the means of each feature $p$ in cluster $k$. Naively, it classically takes $P$ subtractions, $P$ squares, and $P-1$ additions to calculate Euclidean square distance between a feature vector and a centroid vector. This gives rise to each step of the \textit{K}-means algorithm taking time $O(N^{2}PK)$. Allowing ourselves to assume that the states $\ket{\textbf{X}^{i}}$ and the magnitudes $|\textbf{X}^{i}|^2$ are stored in Quantum RAM as would be the case on a true universal quantum computer, we can speed this up exponentially by the quantum method. Whereas classically the $\textit{N}$ different feature vectors $\textbf{X}^{i}$ of dimension $\textit{P}$ require $\textit{P}$ different registers of \textit{n} bits to represent, the same feature vectors can be represented by only $\lceil \log_{2}{\textit{P}}\rceil$ qubits in the computational basis. We can create the state $\ket{\psi}$, where the two vectors we wish to find the distance between are a feature vector and a centroid vector, in time $O(\log{P})$. Then, using a variation of quantum counting and Grover's search algorithm, we can create the state $\ket{\varphi}$ and estimate the normalization constant $Z$ in time $O(\sqrt{\frac{N}{K}})$ \cite{counting, counting1, nielsen_chuang_2000}. From there, we need only apply one \textbf{FREDKIN} gate to calculate the overlap and Euclidean distance. Still we must perform $K\cdot N$ distance calculations per iteration, giving a total complexity of $O(N^{3/2}K^{1/2}\log{P})$. This suggests that our algorithm may be computationally easier than a classical calculation of the Euclidean distance for large datasets. 

\section{Experimental Implementation of Hybrid Clustering Algorithm}
We implement (12) for the calculation of the distance measure between arbitrary-dimensional feature vectors in our clustering algorithm and compare to similar classical algorithms using the open-source IBMQ software for creating and running quantum circuits. We achieved this by utilizing the Python module \textit{Qiskit} developed by IBMQ, coding a modular quantum algorithm for distance calculation and injecting it into a rudimentary \textit{K}-means algorithm. The \textit{Qiskit} module offers the ability to execute circuits on quantum computers operated by IBM built with superconducting transmon qubits and Josephson junctions, as well as on IBM's high performance quantum simulator designed to accurately simulate typical noisy transmons. As in \cite{sim}, we find the experimental error associated with remotely executing circuits on the real devices offered by IBMQ to be too high to extract significant results. Due to this, along with qubit restrictions and long queue times, we executed our algorithm on the simulator to analyze its performance on an unrestricted and less noisy environment, similar to \cite{sim} and \cite{varcircsim}.
Note that while we are using a superconducting processor, there are additional qubit devices that demonstrate promise for machine learning applications such as optical sytems \cite{optics} and trapped-ion processors \cite{ion, ion2}.

\subsection{Clustering}
To test our algorithm we first use the standard \textit{Scikit} function \textit{make\_blobs} to generate gaussian clustered data. We test the algorithm on 4 datasets of 100 5-dimensional feature vectors with different levels of noise and clustering, represented numerically by increasing standard deviations in datapoints from the cluster centroids. We also ran a classical K-means algorithm on the same data for the sake of comparison. Figure 1 shows the clustering of both algorithms on one dataset, while Figure 2 shows accuracy vs. standard deviation values for each algorithm. It is evident that the quantum algorithm performs very similarly to the classical analogue, only suffering a slight performance dropoff for highly noisy data. All simulations were run with \textit{Qiskit Aer} on a Macbook Pro.

\begin{figure}[htbp]
\centerline{\includegraphics[height=5.3cm, width=10.2cm]{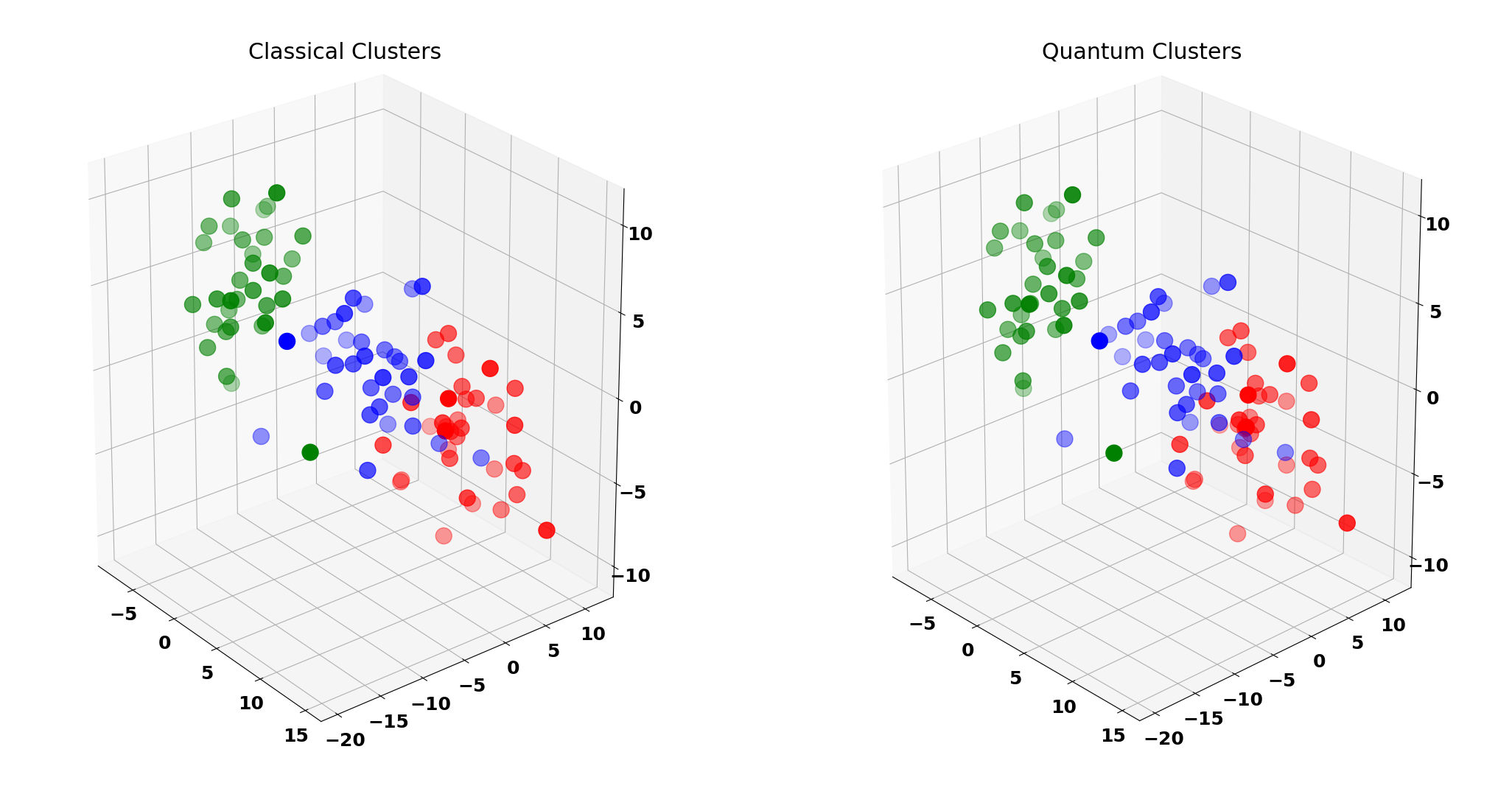}}
\vspace*{13pt}
\caption{\label{fig1}Identified clusters (red, green, blue) by both algorithms in a Gaussian dataset. The dataset was constructed by randomly assigning three cluster centroids with means ranging from [-10, 10]. The datapoints were then randomly generated using a standard deviation of 3.0 away from the cluster centroid, corresponding to a moderately noisy model. Note that the data contains 5 feature dimensions, while only 3 are pictured. Both classical and quantum algorithms perfectly cluster the data, as shown above.}
\end{figure}

\begin{figure}[htbp]
\centerline{\includegraphics[height=5.3cm, width=10.2cm]{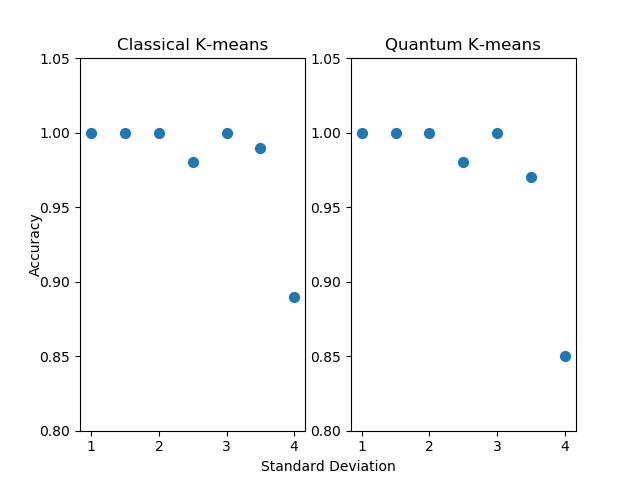}}
\vspace*{13pt}
\caption{\label{fig2}Cluster accuracy vs. standard deviation of data. As the standard deviation of each generated datapoint from its cluster centroid increases, the data becomes more and more noisy. Each algorithm was executed on four datasets of standard deviations ranging from 1.0 to 4.0. Both algorithms perform well below standard deviations of 3.0, and begin to suffer a performance dropoff above that.}
\end{figure}

\subsection{Classification}
Here, we apply our hybrid algorithm to a supervised learning problem on real datasets. We consider a trinary classification problem on the standard Wine and Iris datasets provided by the \textit{Scikit} module for machine learning on Python. We feed the test data into our clustering algorithm, and then compare the generated classes with the true class boundaries. We run the algorithm on each dataset with 30 feature vectors. For the Wine dataset, we find a classification success of 100\% in 5 feature dimensions. For Iris, we find a success rate of 70\% in 4 feature dimensions. In both cases, the algorithm performs similarly to the classical analogue in terms of accuracy. We also executed the algorithm on the Wine dataset for 100 feature vectors in 2 dimensions, finding an accuracy of 98\%, showing the success of the algorithm for higher dimensional feature vectors. Figures 3 and 4 graphically depict one run of the algorithm on each dataset in 2 feature dimensions. We also run our algorithm on the HTRU\_2 dataset \cite{HTRU2}, consisting of a sample of emissions statistics of pulsar candidates. When analyzing 60 feature vectors with 8 dimensions, the classical algorithm achieves an 86.6\% success rate in classifying pulsars, while our algorithm achieves an 83.3\% success rate. Figure 5 depicts one run of the algorithm on HTRU\_2 in two feature dimensions. See section 6 for further discussion and comparison to other algorithms.

\begin{figure}[htbp]
\centerline{\includegraphics[height=6.3cm, width=10.2cm]{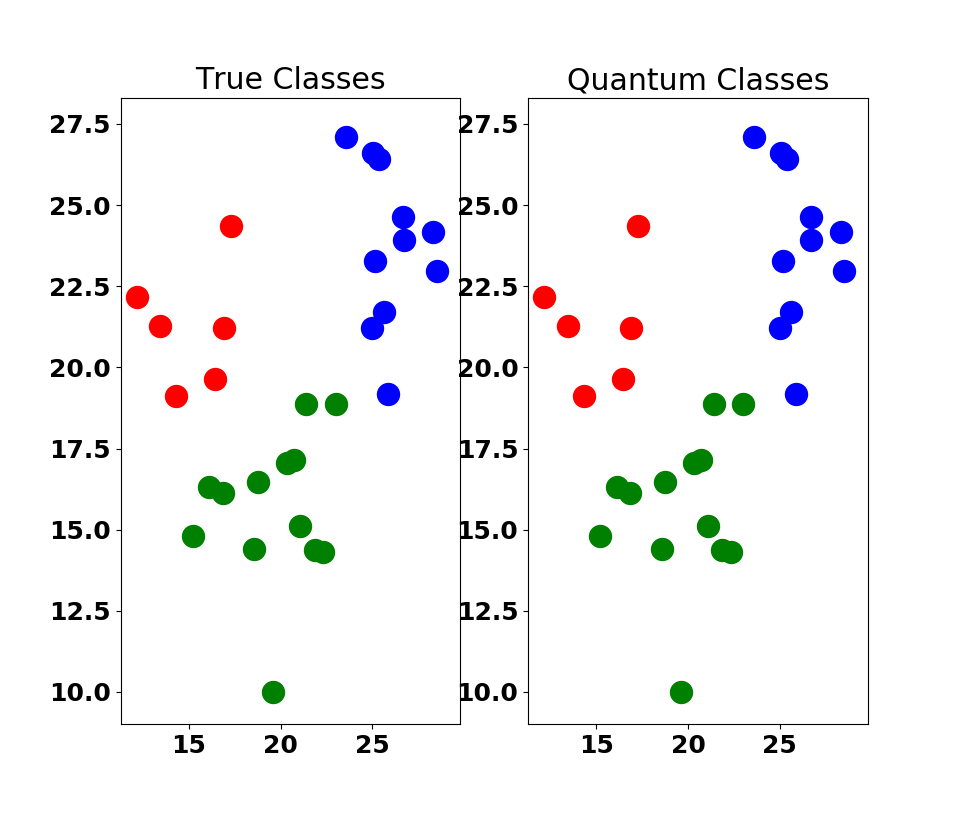}}
\vspace*{13pt}
\caption{\label{fig3}Wine dataset - True and predicted classifications (red, green, blue) with Quantum K-means algorithm. Predicted clusters exactly reflect true boundaries in the data. The natural clustering in the Wine dataset is accurately detected by our algorithm, showing that it tends to detect clustering at a level comparable to similar classical algorithms.}
\end{figure}

\begin{figure}[htbp]
\centerline{\includegraphics[height=6.3cm, width=10.2cm]{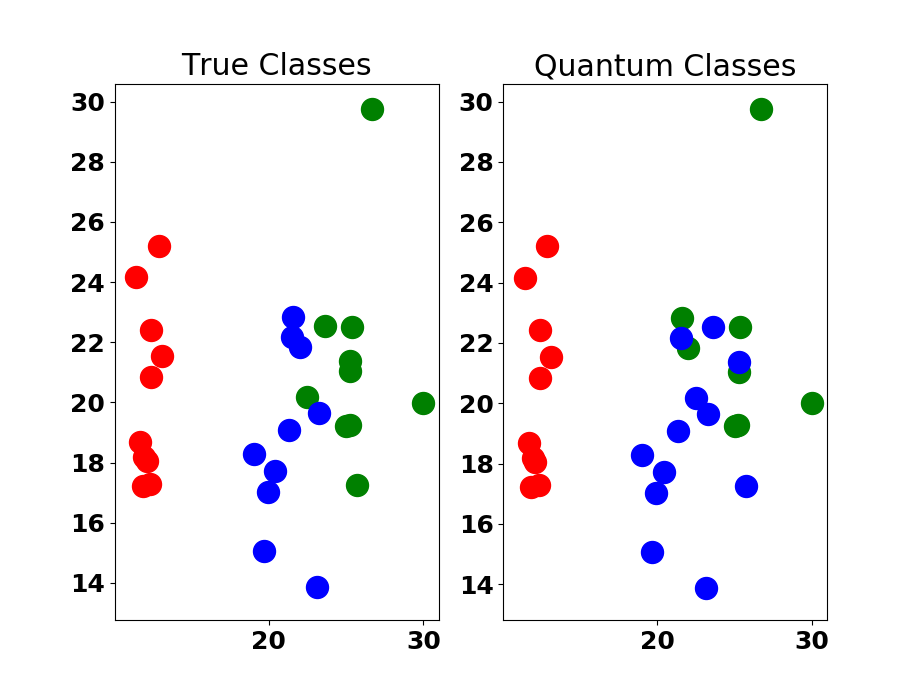}}
\vspace*{13pt}
\caption{\label{fig4}Iris dataset - True and predicted classifications (red, green, blue) with Quantum K-means algorithm. Two of the three predicted classes stray from the true data boundaries. This is because there is little natural separation between the two classes represented on the right side of the graphs, and as such, clustering algorithms do not always perform well at classifying the data.}
\end{figure}

\begin{figure}[htbp]
\centerline{\includegraphics[height=6.3cm, width=10.2cm]{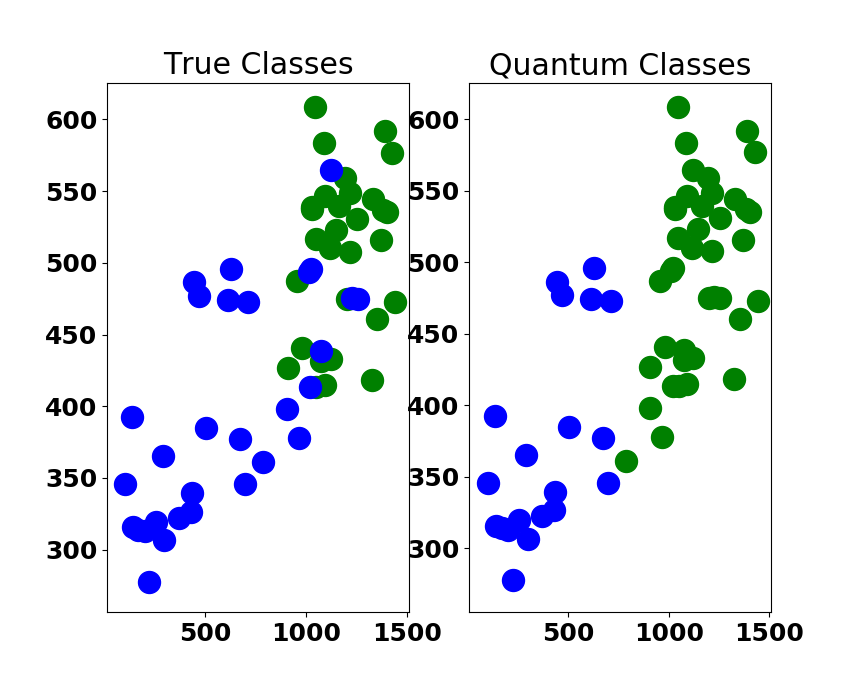}}
\vspace*{13pt}
\caption{\label{fig5}HTRU\_2 dataset - True and predicted classifications (green, blue) with Quantum K-means algorithm. Predicted clusters closely reflect true boundaries in the data. Like in Wine, the natural clustering in the HTRU\_2 dataset is accurately detected by our algorithm.}
\end{figure}

\subsection{Current Limitations}
Some current limitations exist on the IBMQ for applications of this algorithm for calculating quantum distance measures. Our algorithm does not give identical results to the classical analogue due to uncertainty in estimating probability distributions caused by a executing a finite number of circuit runs. For low number of runs (known as \textit{shots} in Qiskit), there is a high variance in calculating the probability needed for (12). We deal with this by running each circuit for 100000 shots, for which we empirically find the standard deviation to be $1.6\%$ of the mean. Also, as stated before, true quantum devices are currently too noisy to extract useful results with this algorithm. As noise mitigation techniques are expectedly improved, this noise will rapidly decrease, allowing for the algorithm to be executed quickly on true quantum devices. 

Additionally, when the optimal cluster centroids for the dataset are located very close together, the algorithm occasionally misclassifies a low number of datapoints near the boundaries due to uncertainty in the distance measure, which can sometimes cause failure of the algorithm to terminate. This is a consequence of the strict condition of termination in the \textit{K}-means algorithm, and slightly easing this condition remedies this problem. 

\section{Quantum Support Vector Machine}

Recently, an algorithm for supervised quantum learning was suggested in \cite{qsvm}. This algorithm used a support vector machine concept on the IBMQ. The support vector machine algorithm attempts to find a separating hyperplane - informally, an $\textit{M}-1$ dimensional generalization of a line - in \textit{M}-dimensional \textit{feature space} (a higher dimensional vector space of nonlinearly transformed feature vectors) from which all of the data points in one class will lie on one side of the hyperplane, and all of the data points in the other class will lie on the other. Let us consider a dataset consisting of \textit{N} different feature vectors $\textbf{X}^{i} = (\textit{X}^{i}_{1}, \textit{X}^{i}_{2},\ldots, \textit{X}^{i}_{P})$. The problem boils down to the convex optimization problem of finding the factors $\alpha^{i}$ such that 
\begin{equation}f(\textbf{X})=\sum_{i=1}^{N}\alpha^{i}y^{i}K(\textbf{X}^{i}, \textbf{X})+b\end{equation} 
is a decision function acting on a datapoint $\textbf{X}$ whose sign correctly classifies it - namely, positive values of $f(\textbf{X})$ correspond to one classification for \textbf{X}, and negative values correspond to the other (see \cite{hastie_tibshirani_friedman_2004, stevens} for an explanation of $y^i$ and $b$). The $\textit{K}(\textbf{X}, \textbf{X}^{i})$ is a positive-definite \textit{kernel} function, equal to an inner product in a high dimensional vector space. 
\begin{equation}\textit{K}(\textbf{X}, \textbf{X}^{i})=\sum_{j=1}^{\infty}\varphi_{j}(\textbf{X}^{i})\varphi(\textbf{X}).
\end{equation} 
Here, $\varphi$ is some non-linear transformation into a higher dimensional vector space. The power of the SVM stems from the fact that the kernel function can be calculated without explicitly calculating $\varphi$. Different kernels give rise to different decision boundaries, but many kernels are classically intractable. However, some can be more efficiently calculated on a quantum computer. 

To achieve a quantum SVM, one must encode the feature vectors $\textbf{X}^{i}$ into quantum states that can be manipulated to compute the desired kernel. According to \cite{qsvm}, define the unitary gate
\begin{equation}
\textbf{U}_{\Phi(\textbf{X}^{i})}=\exp\left(\textbf{i}\sum_{S\subseteq[n]}\phi_{S}(\textbf{X}^{i})\prod_{j\in S}Z_{j}\right),
\end{equation} 
which is diagonal in the Pauli-$Z$ basis, as well as the $n$-qubit gate 
\begin{equation}\textbf{M}_{\Phi(\textbf{X}^{i})}=\textbf{U}_{\Phi(\textbf{X}^{i})}\textbf{H}^{\otimes n}\textbf{U}_{\Phi(\textbf{X}^{i})}\textbf{H}^{\otimes n}
\end{equation} 
where \textbf{H} is the usual Hadamard gate. For example, if $n=2$, the data is encoded through the coefficients $\phi_{S}(\textbf{X}^{i})$, such that 
\begin{equation}\phi_{1}(\textbf{X}^{i})=X_{1}^{i}, \; \phi_{2}(\textbf{X}^{i})=X_{2}^{i}, \; \phi_{\{1, 2\}}(\textbf{X}^{i})=(\pi-X_{1}^{i})(\pi-X_{2}^{i})
\end{equation} 
Define the kernel as 
\begin{equation}K(\textbf{X}^{i}, \textbf{X}^{j})=|\braket{\Phi(\textbf{X}^{i})|\Phi(\textbf{X}^{j})}|^2 = |\braket{0^n|\textbf{M}^{\dag}_{\Phi(\textbf{X}^{i})}\textbf{M}_{\Phi(\textbf{X}^{j})}|0^n}|^2. 
\end{equation} 
This kernel is thought to be classically intractable. However, it can be easily calculated by applying the gate $\textbf{M}_{\Phi(\textbf{X}^{j})}$ followed by the gate $\textbf{M}^{\dag}_{\Phi(\textbf{X}^{i})}$ to an initial state $\ket{0}^{n}$and experimentally calculating the frequency of getting the zero string $0^n$ as a result from the circuit (\cite{qsvm}).

\begin{table}
\centering
\begin{tabular}{|c|c|}
\hline
Algorithm&Accuracy\\
\hline
Classical SVM&96.7\%\\
\hline
Quantum SVM by Havlicek et al.&63.3\%\\
\hline
Classical K-means&88.7\%\\
\hline
Quantum K-means&96.7\%\\
\hline
\end{tabular}
\caption{Trinary Classification on Wine Dataset. Each row depicts one algorithm, showing the average accuracy of all 5 trials. As the Wine dataset contains natural clustering, both classical and quantum K-means clustering algorithms perform very well at classifying the data. The quantum clustering algorithm performs better than its classical analogue. }
\end{table}

\begin{table}
\centering
\begin{tabular}{|c|c|}
\hline
Algorithm&Accuracy\\
\hline
Classical SVM&93.3\%\\
\hline
Quantum SVM by Havlicek et al.&80.0\%\\
\hline
Classical K-means&72.7\%\\
\hline
Quantum K-means&75.3\%\\
\hline
\end{tabular}
\caption{Trinary Classification on Iris Dataset. Each row depicts one algorithm, showing the average accuracy of all 5 trials. As the Iris dataset does not contain much natural clustering, both classical and quantum K-means clustering algorithms suffer in performance at classifying the data. Still, the quantum clustering algorithm performs better than its classical analogue.}
\end{table}

\begin{table}
\centering
\begin{tabular}{|c|c|}
\hline
Algorithm&Accuracy\\
\hline
Classical SVM&64.3\%\\
\hline
Quantum SVM by Havlicek et al.&49.3\%\\
\hline
Classical K-means&85.0\%\\
\hline
Quantum K-means&83.3\%\\
\hline
\end{tabular}
\caption{Binary Classification on HTRU\_2 Dataset. Each row depicts one algorithm, showing the average accuracy of all 5 trials. The natural clustering in the dataset leads to the K-means algorithms performing well. Both SVM algorithms suffer greatly on this dataset.}
\end{table}

\section{Comparison of the Algorithms}
In this section, we further compare our quantum machine learning to \cite{qsvm} on the IBMQ with real datasets Wine and Iris. While \textit{K}-means clustering and the support vector machine belong to different classes of machine learning algorithms, with the former being a clustering algorithm and the latter being a supervised binary classification algorithm, we can compare them under certain constraints. For the \textit{K}-means clustering algorithms, we input the unlabeled test data and then compare the generated clusters with the true labels of the data. Thereafter, we calculate the accuracy by calculating the percentage of correctly classified features. For the Support Vector Machine, we utilize the \textit{One Against Rest} multiclass extension provided by \textit{Qiskit} to extend the SVM to more than two classes. The \textit{One Against Rest} extension constructs a number of SVMs equal to the number of classes, each of which compare one class against all the others. See \cite{multiclass} for more details. We include a classical \textit{K}-means clustering algorithm as well as a classical RBF kernel SVM for comparison. We run five experiments on each classifier, with 2 feature dimensions and 30 test inputs, as well as 30 training inputs for the SVMs. Note that the clustering algorithms are not provided with any training vectors and corresponding labels, while the SVMs are. The results of the simulations are shown in Tables 1 and 2. Additionally, we run all four algorithms on the HTRU\_2 binary classification dataset with the same dimension and number of inputs. The results are shown in Table 3.

On the Wine dataset, the hybrid clustering algorithm performs comparably to the classical SVM, with the quantum SVM of \cite{qsvm} falling behind in terms of accuracy. For the Iris dataset, which contains little natural clustering between two of the three classes, the SVM algorithms perform better as their kernels allows them to classify non-linear boundaries. Yet, the quantum \textit{K}-means clustering algorithm is more accurate than its classical counterpart on both datasets. Finally, for the HTRU\_2 dataset, the K-means algorithms perform significantly better than the SVMs. Like in Wine, the K-means algorithms identify the natural clustering in the dataset well. The SVM algorithms' performance may be suffering due to the low number of training vectors. 

\section{Conclusions}
In this paper we introduced a quantum algorithm for \textit{K}-means clustering, based on the standard classical clustering algorithm and a novel quantum technique for the computation of Euclidean distance. We find that the algorithm very closely matches the classical \textit{K}-means algorithm in accuracy on clustering and classification problems, while showing the potential of being significantly computationally cheaper when executed on a true quantum computer. This marks another step in the field of quantum machine learning in designing and implementing quantum algorithms which are as accurate and potentially faster than their classical counterparts.

\end{document}